\documentstyle[aps,epsf,12pt]{revtex}
\tightenlines
\begin{document}
\pagestyle{empty}
\title{Periodic Instantons in $SU(2)$ Yang-Mills-Higgs Theory}
\vskip1cm
\author{G.~F.~Bonini}
\address{Institut f\"ur Theoretische Physik \\ 
Universit\"at Heidelberg \\ 
D-69120 Heidelberg, Germany\\ 
bonini@thphys.uni-heidelberg.de}

\author{S.~Habib, E.~Mottola}
\address{Theoretical Division, T-8 \\ 
Los Alamos National Laboratory \\  
Los Alamos, NM 87545 USA \\ 
habib@lanl.gov, emil@lanl.gov}

\author{C.~Rebbi}

\address{Department of Physics, Boston University \\
Boston, MA 02215 USA\\ rebbi@bu.edu}

\author{R.~Singleton}

\address{Department of Physics, University of Washington\\
Seattle WA 98195 USA\\ 
bobs@terrapin.phys.washington.edu}

\author{P.~G.~Tinyakov}

\address{Institute for Nuclear Research \\ 
60 October Anniversary Pr. 7a\\
Moscow 117312, Russia\\ peter@ms2.inr.ac.ru}
\vskip1cm

\maketitle
\begin{abstract}
The properties of periodic instanton solutions of
the classical $SU(2)$ gauge theory with a Higgs doublet field are described
analytically at low energies, and found numerically for all energies up to and
beyond the sphaleron energy. Interesting new classes of bifurcating
complex periodic instanton solutions to the Yang-Mills-Higgs equations
are described.
\end{abstract}

\vskip1cm
\noindent BUHEP-99-10  \\
\noindent HD-THEP-99-13 \\
\noindent UW/PT 999-08 \\
\noindent LA-UR-99-2295\\

\setcounter{page}{0}
\thispagestyle{empty}

\pagestyle{empty}
\newpage
\pagestyle{plain}
\pagenumbering{arabic}

\noindent {\em Introduction.}
Anomalous baryon and lepton number violating processes in the electroweak
theory are dominated in the semiclassical weak coupling limit,
$\alpha_{_W} \rightarrow 0$, by field configurations which solve the
classical Euclidean Euler-Lagrange equations. At zero temperature
and energy the classical solutions that contribute to anomalous winding number
transitions are the familiar BPST instantons/anti-instantons~\cite{BPST}.
The rate of these vacuum tunneling transitions is exponentially suppressed
$\sim \exp (-2 S_I)$, where $2S_I = {4\pi/\alpha_{_W}}$ is the
Euclidean action of a widely separated instanton (I)/anti-instanton
($\bar I$) pair. At temperatures much higher than $M_W$, the transitions are
classical thermal activation transitions over the potential barrier
between vacua, with a Boltzmann rate $\sim \exp (-E_s/k_BT)$ controlled
by the energy $E_s \sim 4 M_W/\alpha_W$ of a certain unstable
classical stationary field configuration called the sphaleron~\cite{Man}.

At intermediate temperatures, or at finite energy (not necessarily
arranged in thermal equilibrium), very little is known about the
rate of anomalous transitions between states of different winding
number. The first step in studying these transitions is to
find the classical solutions which dominate the semiclassical
rate, and calculate their action. This involves solving the
classical nonlinear field equations in Euclidean time.
At low energies one can construct solutions of the classical
Euclidean Yang-Mills-Higgs equations consisting of
periodic chains of $I\bar I$ pairs arrayed along the imaginary
time axis. The action of these periodic instanton solutions
can be expressed as a power series in $\left(E/E_s\right)^{2/3}$
for small $E/E_s$~\cite{KRT}. This perturbative treatment of low energy
periodic instanton solutions can be recast as an expansion in powers
of $(M_W\beta)^2$ where $\beta$ is the period of the solution.
For larger energies or periods the solutions can be found numerically.
In this Letter we describe the qualitative properties of and
present numerical results for these periodic instanton solutions of the $SU(2)$
Yang-Mills-Higgs equations, {\it i.e.}~the bosonic sector of the standard
electroweak theory with $\theta_W = 0$. An unexpectedly rich structure of
bifurcating periodic instanton solutions, both real and complex has been
found, whose physical consequences for B and L violating transitions in the
electroweak theory remains to be more fully investigated.\\

\noindent {\em Periodic Instantons at Low Energy.}
As a simple example of a periodic potential with periodic instanton
solutions consider the pendulum potential,
\begin{equation}
V(q) = \omega^2 ( 1 - \cos q )\,.
\end{equation}
The zero energy instanton which interpolates between the
vacuum states at $q = 0$ and $q = 2 \pi$ is the kink
configuration, $q_I(\tau)$, given by
\begin{equation}
\cos \left({q_I(\tau)\over 2}\right) = - {\rm tanh} (\omega\tau)\,
\end{equation}
which solves the classical Euclidean equations $\ddot q_I = V'(q_I)$
and has action $S_I = 8 \omega$. The anti-instanton solution is
$q_{\bar I}(\tau) = q_I(-\tau)$ with the same action. Consider
now the widely separated $I-\bar I$ pair configuration,
\begin{equation}
q_{I\bar I}(\tau) = q_I(\tau) + q_{\bar I}(\tau - \bar\tau) - 2\pi\,
\end{equation}
with $\bar\tau \gg 1/\omega$. The action of this
configuration can be computed to first order in the interaction
between the pair, with the result
\begin{equation}
S [ q_{I\bar I}] = 2 S_I -32 \,\omega \,e^{-\omega \bar\tau} + {\cal O}
\left(e^{-2\omega\bar\tau}\right)\,.
\end{equation}
The negative sign reflects the attractive interaction between
the $I$ and $\bar I$.

We now consider a periodic arrangement of $I$ and $\bar I$ at
equal intervals along the imaginary time axis, with period
$\beta$. This means that the separation between nearest neighbor
$I$ and $\bar I$ is $\bar\tau = \beta/2$. The attractive
force between neighbors can now exactly balance and yield
an extremum of the action, the periodic instanton solution.
Since there are two nearest neighbor $I\bar I$ interactions per
period we expect the action per period of this solution to be
\begin{equation}
S(\beta) = 16 \,\omega - 64 \,\omega e^{-\omega\beta/ 2} +
{\cal O} \left(e^{-\omega\beta}\right)\,,
\end{equation}
in the limit of large $\omega\beta$. Since
\begin{equation}
E(\beta) = {d S(\beta)\over d\beta} = 32 \,\omega^2
e^{-\omega\beta/ 2} + \dots
\end{equation}
large $\beta$ corresponds to low energy. In the one dimensional
pendulum example the exact periodic instanton solution with this
action and energy are easily found explicitly in terms of elliptic
functions by simple quadrature. Because of the attractive
interaction between the $I$ and $\bar I$ along the chain it is
clear that there is a single negative mode of the second order
fluctuation operator, $-\partial_{\tau}^2 + V''(q(\tau))$, around this
periodic instanton solution, a fact
that is also reflected by the second derivative of the action,
\begin{equation}
{d^2 S(\beta)\over d\beta^2} = {d E(\beta)\over d\beta} =
-16\,\omega^2 e^{-\omega\beta/ 2} + \dots \ < 0\,.
\label{sec}
\end{equation}
The monotonic decrease of period $\beta$ with increasing energy
persists up to $E = E_s = 2\omega^2$, where the curve of $S(\beta)$
vs.~$\beta$ of the periodic instanton becomes tangent to the
constant sphaleron solution, corresponding in this simple model
with the unstable static configuration $q_{_s} = \pi$. This occurs at
$\beta = \beta_- = 2\pi/\omega$ equal to the period of oscillation
in the inverted potential at $q = q_{_s}$. At this $\beta$ the action
of the periodic instanton is $E_s \beta_- = 4\pi\omega < 16 \omega$,
reflecting the fact that the action is monotonically decreasing
as $\beta$ ranges from $\infty$ down to $\beta_-$, and as $E$ increases
from $0$ to $E_s$.

Beyond the point where the periodic instanton and sphaleron solutions
merge, a complex solution bifurcates from the sphaleron and continues
with real decreasing action. This may be understood by the amplitude
of the zero mode at $\beta = \beta_-$ turning from real to pure
imaginary as $\beta$ is decreased through the critical value.
The generic behavior described here is what we call
type~(I) behavior of the periodic instanton solutions, for
which the monotonic negative sign in the first derivative of
the action in (\ref{sec}) and $E_s \beta_- < 2S_I$
are characteristic.

A different pattern is possible when the instanton has additional
zero modes, and therefore additional parameters enter the
description. Such is the case in field theory models with
exact or softly broken conformal invariance. We have studied this case
in some detail in the $O(3)$ nonlinear sigma model in two dimensions,
softly broken by a mass term~\cite{Othree}. This model shares many features
with the bosonic sector of the $SU(2)$ electroweak theory. Although
there is no isolated single $I$ or $\bar I$ solution in the
broken theory, due to Derrick's theorem (which tells us that
zero scale size $\rho \rightarrow 0$ has minimum action), a periodic
instanton solution does exist in which the scale size $\rho$ is
adjusted to a certain value as a function of period $\beta$. At
this value the attractive interaction between $I$ and $\bar I$
exactly balances the tendency of each individual $I$ or $\bar I$
to collapse to zero size. It is again possible to understand
this at low energies by first finding the two-body interaction between
well isolated $I$ and $\bar I$, and then arranging them periodically
along the imaginary time axis, calculating the change in the action
per period from that of a single $I \bar I$ pair due to the sum
of the first order interactions between them. In this calculation
the scale size $\rho$ can be treated as a variational parameter
with the value on the solution $\rho (\beta)$ determined by
extremizing $S(\beta, \rho)$ with respect to $\rho$. Substituting
into $S$ then gives $S(\beta)$ on a low energy periodic
instanton solution. The resultant behavior depends upon the
curvature of $S(\beta)$, which is of {\it opposite} sign relative
to the type~(I) models, and we will consequently refer to this
case as type~(II),~{\it i.e.}
\begin{equation}
{d^2 S(\beta)\over d\beta^2} = {d E(\beta)\over d\beta} > 0\,,
\qquad {\rm type \ (II)}.
\label{two}
\end{equation}

The periodic instanton solutions in this
case have two negative modes rather than just one, with the
second negative mode corresponding to variation of the scale
size $\rho$ away from its extremal value $\rho (\beta)$. Such
periodic instanton solutions do {\it not} contribute to
thermal winding number processes at finite temperature.
However, they can contribute to anomalous finite-energy
non-thermal transitions.

Because of the existence of the conformal mode in $SU(2)$
BPST instantons, we would expect the Yang-Mills-Higgs theory
to behave qualitatively similar to the $O(3)$ sigma model, and
to also be of type (II). Indeed for low energies we
can show that this is exactly what happens. The
action for an isolated pure $SU(2)$ $I \bar I$ pair
with scale size $\rho$ separated by distance $\bar\tau$ is
\begin{equation}
S_{I\bar I} = 2 S_{I} - {96 \pi^2 \rho^4 \over g^2 \bar\tau^4} +
{\cal O} \left({\rho^6\over g^2\bar\tau^6}\right)\,,
\end{equation}
with $S_I = 8\pi^2/g^2 = 2\pi/\alpha$ the single instanton
action and the second term the well-known dipole-dipole attractive
interaction between the $I$ and $\bar I$ aligned in an internal
$SU(2)$ direction. When the periodic chain of $I$ and $\bar I$
separated by $(n + {1\over 2})\beta$ is constructed, this
leads to the total interaction,
\begin{equation}
S_{\rm int} = - {96 \pi^2 \rho^4 \over g^2
\beta^4}\sum_{n=-\infty}^{\infty}
{1\over (n + {1\over 2})^4} =  -{4\pi\over \alpha}
\left({2\pi^4\rho^4\over \beta^4}\right)\,.
\end{equation}
When the $SU(2)$ doublet Higgs field is added to the action it
can be solved for at leading order in the $I$ or $\bar I$
background and gives a contribution,
\begin{equation}
S_{\rm Higgs} = {4\pi\over \alpha}\left({M_W^2 \rho^2\over 2}
\right) + {\cal O} \left({M_W^2\rho^4\over g^2\beta^2}\right)\,,
\end{equation}
which was first calculated by `t Hooft~\cite{tH}.

This positive contribution
expresses the fact that Derrick's theorem drives the single
isolated $I$ or $\bar I$ scale size to zero $\rho$; however, at finite
$\beta$ this is opposed by the dipole-dipole interaction
$S_{\rm int}$, and the variational action
\begin{equation}
S(\beta, \rho) = {4\pi\over \alpha}\left[ 1 + {M_W^2 \rho^2\over 2}
-{2\pi^4\rho^4\over \beta^4} + {\cal O} \left({M_W^2\rho^4\over \beta^2},
{\rho^6 \over \beta^6}\right)\right]
\end{equation}
has an nontrivial extremum at $\rho = \rho (\beta)$ when
$\partial S/\partial \rho = 0$, or
\begin{equation}
 \rho = {\sqrt 2 \over M_W} \left[
 x^2 + {\cal O} \left( x^4 \right) \, \right]\,,
\end{equation}
denoting by $x$ the expansion parameter $M_W \beta/  (2\pi)$.
This stationarity condition implies that the various next-to-leading
contributions to the action from both the gauge and Higgs fields are
all of order $x^6$. Hence the periodic instanton action,
\begin{equation}
S = {4\pi\over \alpha}\left[1 + {1\over 2}
x^4 + {\cal O}
\left(x^6\right)\right]
\label{Spert}
\end{equation}
for small $M_W\beta$, and
\begin{equation}
E = {d S(\beta)\over d\beta} = {4M_W\over\alpha}
x^3\left[1 +
{\cal O}\left(x^2\right)\right]
\label{Epert}
\end{equation}
is an increasing function of period $\beta$ for small $M_W\beta$.
The second derivative of $S(\beta)$ is also clearly positive.
After a rather elaborate calculation~\cite{BHMRST}, one can
evaluate the coefficients of the next to leading terms in the
perturbative expansion. One thus finds
\begin{eqnarray}
S &=& {4\pi\over \alpha}\left[1 + {1\over 2}
x^4 + {4 \over 3} x^6  + {\cal O}
\left(x^8\right)\right]
\label{Spert2} \\
E &=& {4M_W\over\alpha}
x^3\left[1 + 4x^2 +
{\cal O}\left(x^4\right)\right] \ .
\label{Epert2}
\end{eqnarray}

Although the $SU(2)$ Higgs theory starts out at low energy and small
$\beta$ behaving like the type (II) case, there is an additional independent
parameter in the $4D$ gauge theory, namely the quartic Higgs self-coupling
$\lambda$, or equivalently the Higgs mass, $M_H$. Thus, we cannot
preclude a more complicated behavior at larger energies, depending
on the value of $M_H/M_W$. This is indeed what we have found in
our numerical study.\\

\noindent {\em Periodic Instantons at Finite Energy in the
$SU(2)$-Higgs Theory.}
We consider the $SU(2)$ gauge theory with a doublet Higgs field
in $4D$ Euclidean space with the action,
\begin{eqnarray}
S &=& \frac{1}{g^2}\int d^4x ~ \left\{{1 \over 2}
  {\rm Tr}\,(F_{\mu \nu} F_{\mu \nu}) + (D_{\mu}
  \Phi)^\dagger (D_{\mu} \Phi)  + \frac{\lambda}
  {g^2} \left(\Phi^\dagger \Phi - \frac{g^2v^2}
  {2} \right)^2\right\} \,,\nonumber\\
&&F_{\mu\nu} = \partial_\mu A_\nu - \partial_\nu A_\mu - i [A_\mu,A_\nu]\,,
\quad D_\mu \Phi = (\partial_\mu - i A_\mu) \Phi\,,
\label{fourAction}
\end{eqnarray}
and $A_\mu = A^a_\mu\sigma^a/2$.
The corresponding classical Euler-Lagrange equations are
\begin{eqnarray}
  D_\mu F_{\mu\nu} + i (D_\nu \Phi^\dagger)\times\Phi -
  i \Phi^{\dagger}\times (D_\nu \Phi) &=& 0
\nonumber\\
 \left[-D^2 + \frac{2\lambda}{g^2}\left(
  \Phi^\dagger \Phi - \frac{g^2 v^2}{2}\right)
 \right]\Phi &=& 0 \,,
\label{fourEqs}
\end{eqnarray}
where the covariant derivative acting on the $A$-field
in the adjoint representation is \hbox{$D_\mu A_\nu =
\partial_\mu A_\nu - i\, [A_\mu,A_\nu]$}, and the $\times$
denotes the outer product of the two spinors. We use the
standard conventions for the Higgs vacuum expectation value
$v$ and the self-coupling $\lambda$ in which the $W$- and
and Higgs-masses are $M_W = \frac{1}{2} g v$ and
$M_H = \sqrt{2\lambda} v$ respectively.

The spherical {\it Ansatz} is given by expressing the gauge
field $A_\mu$ and the Higgs field $\Phi$ in terms of six
real functions  $a_0\, ,\,a_1\, , \, \alpha\, , \, \gamma\, , \,
u\ {\rm and}\ w\ {\rm of}\ r\ {\rm and}\ \tau$:
\begin{eqnarray}
  A_0({\bf x}, \tau) &=& \frac{1}{2} \, a_0(r,\tau)\,
  {\vec\sigma}\cdot{\bf \hat x}
\nonumber\\
  A_i({\bf x}, \tau) &=& \frac{1}{2} \, \big[a_1(r,\tau)\,
  {\vec\sigma}\,\cdot{\bf\hat x}\,\hat
  x_i+\frac{\alpha(r,\tau)}{r}\,(\sigma_i-{\vec\sigma}
  \cdot{\bf\hat x}\,\hat x_i)+\frac{\gamma(r,\tau)}{r}\,
  \epsilon_{ijk}\,\hat x_j\,\sigma_k\big]\nonumber\\
  \Phi({\bf x}, \tau) &=& \sqrt{2}\,M_W  \,
  [u(r,\tau) + i w(r,\tau)\,
  {\vec\sigma}\cdot{\bf \hat x}]\, \hat\zeta \ ,
\label{SphAn}
\end{eqnarray}
where $\hat\zeta$ is an arbitrary unit two-component spinor,
${\bf \hat x}$ is the unit three-vector in the radial
spatial direction, and ${\vec\sigma}$ is a three-vector of
Pauli matrices.

Upon substituting (\ref{SphAn}) into the action
(\ref{fourAction}) one finds~\cite{RYW}
\begin{eqnarray}
S &=&\frac{4\pi}{g^2}  \int d\tau\int^\infty_0dr \, \bigg[\frac{1}{4}
r^2f_{\mu\nu}f_{\mu\nu}+(\bar D_\mu \bar \chi)D_\mu \chi
+ r^2 (\bar D_\mu \bar\phi)D_\mu\phi
+\frac{1}{2 r^2}\left( ~\bar\chi\chi-1\right)^2\nonumber\\
& &+\frac{1}{2}(\bar\chi\chi + 1)\bar\phi\phi +  {\rm Re}(i \bar\chi \phi^2)
  +\frac{\lambda}{g^2}  \, r^2 \, \left(\bar\phi\phi- 2M_W^2\right)^2 \bigg]
\ , \label{effAction}
\end{eqnarray}
where the indices now run over $0$ and $1$ and
\begin{eqnarray}
  f_{\mu\nu}&=&\partial_\mu a_\nu - \partial_\nu a_\mu\,,\nonumber\\
  \chi =\alpha+i\Big(\gamma-1\Big)\,&,& \quad
  \bar\chi = \alpha - i\Big(\gamma-1\Big)\,,\nonumber\\
  \phi = \sqrt{2}\,M_W\,(u + i w)\,&,& \quad
  \bar\phi = \sqrt{2}\,M_W\,(u - i w)\,,\nonumber\\
  D_\mu\chi = (\partial_\mu-i  \, a_\mu)\chi\,&,&\quad
  \bar D_\mu\bar\chi = (\partial_\mu + i  \, a_\mu)\bar\chi\,,\nonumber\\
  D_\mu \phi= (\partial_\mu- \frac{i}{2}  \, a_\mu)\phi\,&,& \quad
  \bar D_\mu \bar\phi= (\partial_\mu + \frac{i}{2}  \, a_\mu)\bar\phi\ .
\label{defns}
\end{eqnarray}
The equations of motion for the reduced theory are
\begin{eqnarray}
&& -\partial_\mu(r^2f_{\mu\nu})=
  i\left[(\bar D_\nu \bar\chi)\chi-\bar\chi D_\nu\chi \right]+
  \frac{i}{2}\,   \, r^2 \left[(\bar D_\nu \bar\phi)\phi-
  \bar\phi D_\nu\phi\right]\,,\nonumber\\
&&  \left[-D_{\mu}D_{\mu}+\frac{1}{r^2}(\bar\chi\chi-1) + \frac{1}{2}\,
  \bar\phi\phi ~\right]\chi=
  -\frac{i}{2}\,  \, \phi^2\,,\nonumber\\
&&  \left[-D_\mu (r^2 D_\mu) + \frac{1}{2}(\bar\chi\chi+1) +
  \frac{2\lambda}{g^2}\,r^2 \left(\bar\phi\phi-
  2M_W^2\right)\right] \phi= i \, \chi \bar\phi \ .
\label{Eqsom}
\end{eqnarray}
\noindent
Note that the overbar on $\phi$, $\chi$ and $D_{\mu}$ denotes
changing $i \to -i$ in the definitions (\ref{defns}) above,
which is the same as complex conjugation {\it only} if
the six fields $a_{\mu}, \alpha, \gamma, u$ and $w$ are real.
These equations can be obtained by either imposing
the the spherical {\it Ansatz} (\ref{SphAn}) on the four dimensional
equations (\ref{fourEqs}), or by varying the action
(\ref{effAction}) directly.
\vfil\eject
The spherical {\it Ansatz} (\ref{SphAn}) has a residual $U(1)$
gauge invariance under the $U(1)$ gauge transformation,
\begin{eqnarray}
a_\mu &&\to a_\mu + \partial_\mu \Omega\,,\nonumber\\
\chi \, &&\to e^{i \Omega} \chi\,,\nonumber\\
\phi \, &&\to e^{i \Omega/2} \phi  \ ,
\label{gaugexform}
\end{eqnarray}
The complex scalar fields $\chi$ and $\phi$ have
$U(1)$ charges of $1$ and $1/2$ respectively, $a_{\mu}$ is the $U(1)$
gauge field, $f_{\mu\nu}$ is the field strength, and $D_{\mu}$ is the
covariant derivative. The residual $U(1)$ gauge invariance must
be fixed for numerical solution of the equations. In our numerical
work we chose the temporal gauge $a_0 = 0$. The remaining time independent
gauge freedom is fixed by a boundary condition at the quarter period
time slice $\tau = \beta/4$ to be specified below. In the $a_0 = 0$
gauge the $\nu =0$ component of the first of Eqs.~(\ref{Eqsom}),
{\it i.e.}~the Gauss law constraint, must be imposed on the initial $\tau =0$
surface, whereupon it will be satisfied for all $\tau$.

The action of the various discrete symmetries, C, P, and T on the two
dimensional fields follows directly from the spherical {\it Ansatz}
(\ref{SphAn}). In addition to these symmetries, we may consider
the two dimensional reflection symmetry $R: \phi \to -\phi$. 
We can employ these discrete symmetries to help us select the
appropriate boundary conditions for the periodic instanton solution. 
The $CPR$ even fields are $\gamma$ and $w$, whereas the other
four fields are $CPR$ odd. Since we are searching for a periodic
solution which returns to itself with period $\beta$, the time
derivatives must reverse sign in the second half period relative
to the first. This means that we should require the boundary
condition that the $\tau$ derivatives of all remaining five
functions in $a_0$ gauge vanish at $\tau = 0$ and $\tau = \beta/2$. 
This corresponds to an instanton at $\beta/4$ and an anti-instanton 
at $3\beta/4$ in the low energy limit. At the time slice $\tau =\beta/4$ 
the fields are sphaleron-like. In a gauge where the four dimensional 
fields are regular at the origin, the sphaleron is a $CPR$ even 
configuration, and therefore the $CPR$ odd fields change sign 
while the $CPR$ even fields reach a maximum at $\tau = \beta/4$. 
Hence we actually require the solution only on the quarter interval 
$[0,\beta/4]$, if we specify the boundary conditions,
\begin{eqnarray}
\dot a_1 = \dot \alpha = \dot \gamma = \dot u = \dot w = 0\,,
\qquad \tau &= &0\,;\nonumber\\
a_1 = \alpha = u = 0 = \dot\gamma = \dot w\,,\qquad \tau &=& {\beta\over 4}
\,.
\end{eqnarray}
These boundary conditions eliminate the time translational zero mode.

The Euler-Lagrange Eqs.\ref{Eqsom} are also invariant under the two
additional complex discrete transformations:
\begin{eqnarray}
{\cal C}_1:&& \qquad
\phi \rightarrow -\bar\phi\,,\qquad
\bar\phi \rightarrow -\phi\,,\qquad
\chi \rightarrow -\bar\chi\,,\qquad
\bar\chi \rightarrow -\chi\,,\qquad
a \rightarrow -a
\\
{\cal C}_2:&& \qquad
\phi \rightarrow \bar\phi^*\,,\qquad \hskip0.12cm
\bar\phi \rightarrow \phi^*\,,\qquad \hskip0.17cm
\chi \rightarrow \bar\chi^*\,,\qquad \hskip0.17cm
\bar\chi \rightarrow \chi^*\,,\qquad \hskip0.17cm
a \rightarrow a^* \ .
\end{eqnarray}
Under ${\cal C}_1$: $S\rightarrow S$, under ${\cal C}_2$:
$S\rightarrow S^*$ (and similarly for $E$).

For the four dimensional fields to be regular at the origin, and
to approach the vacuum at $r = \infty$ we require the boundary
conditions,
\begin{eqnarray}
\alpha = \gamma =  w = 0 = a_1' = u'\,,\qquad r &= &0\,;\nonumber\\
a_1 = \alpha = u = 0, \qquad {\rm but}\qquad \gamma = 2,\  w = 1\,,\qquad r &=&
\infty\, ,
\end{eqnarray}
where the last condition is necessary for a nonzero winding
number, and agrees with the sphaleron boundary condition at 
$r =\infty$ on the $\tau = \beta/4$ slice. The boundary conditions
on $a_1$ at the origin and infinity are gauge choices, which
completely eliminate the time independent gauge freedom
in temporal gauge $a_0=0$.\\

\noindent {\em Numerical Results.} With a well-defined elliptic
boundary value problem standard numerical methods may be applied.
Here we only outline our computational procedure, the full
details of which will be presented in a separate
publication~\cite{BHMRST}.
We discretized the Euler-Lagrange equations in space and time
by using link variables for the gauge degrees of freedom, thus preserving exact
gauge invariance under space dependent gauge transformations (consistent
with our choice of the $a_0=0$ gauge). The code allows for non-uniform grids in
both space and time to better approximate the fields in the regions of fastest
variation. We also allowed for the analytic continuation of the solutions into
complex valued functions by using complex variables to represent the fields
$a_1, \alpha, \gamma, u, w$ and by discretizing the equations in a manner
compatible with analytic continuation. The coordinates $(r, \tau)$ are
maintained real.

The equations were solved by the Newton-Raphson technique.
Denoting a definite trial configuration of the fields by
$f_i$, where $i$ stands for the discretized two dimensional
lattice
point $(r, \tau)$ as well as the various field components
themselves, we may calculate the gauge invariant discretized
action functional $S[f]$, the discretized first variation,
$\partial S / \partial f_i$, and the second order fluctuation
operator $\partial^2 S / \partial f_i \partial f_{i'}$ on the
trial configuration. Since the first variation must vanish on the
solution, one can find the first order correction to the
configuration, $\delta f$, by solving the linear equations,
\begin{equation}
\sum_{i'}{\partial^2 S \over\partial f_i \partial f_{i'}}\
\delta f_{i'}  + {\partial S \over \partial f_i} = 0\,.
\end{equation}
Adding $\delta f$ to $f$ yields a corrected trial configuration,
and the process may now be iterated. Clearly, all gauge and translational zero
modes must be removed from the second order variation by the gauge fixing and
boundary conditions in order for the inverse of $\partial^2 S /
\partial f_i \partial f_{i'}$ to exist and the procedure to be well
defined.

The algorithm converges quite rapidly, the error decreasing quadratically
with the number of complete Newton-Raphson iterations. The
most time and memory consuming step is the inversion of the second order
fluctuation operator. With grids consisting of as many as $128\times
128$ points (or more) and 5 complex valued fields per point, a direct
solution of the above system of linear equations is prohibitive. Because the
equations of motion are local, the matrix to be inverted is a sparse banded
matrix, and it is much more efficient to use the method of
eliminating alternate time slices instead of a direct inversion.
This effectively reduces the dimensionality of the linear
system one must invert to the size of the space grid only, times the 5
components of the fields. Quite modest lattices ($64 \times 64$) are
sufficient for accuracy of order one percent, except in the small $\beta$, low
energy region, where the larger, adaptive lattices ($128 \times 128$ or
greater) were used. From this solution at a given $\beta$ and $\lambda$ other
solutions were found by using the previous one as a trial
configuration for the new values of the parameters, changing the
values of period and $\lambda$ in small increments.

Our results for the dependence of the action on the period are shown
in Figs.~1 for two different values of
the quartic Higgs coupling, $\lambda = 0.7g^2$ and $\lambda = 3.6g^2$,
respectively. They clearly exhibit a pattern of bifurcations.

\begin{figure}
\epsfxsize=12cm
\centerline{\epsfbox{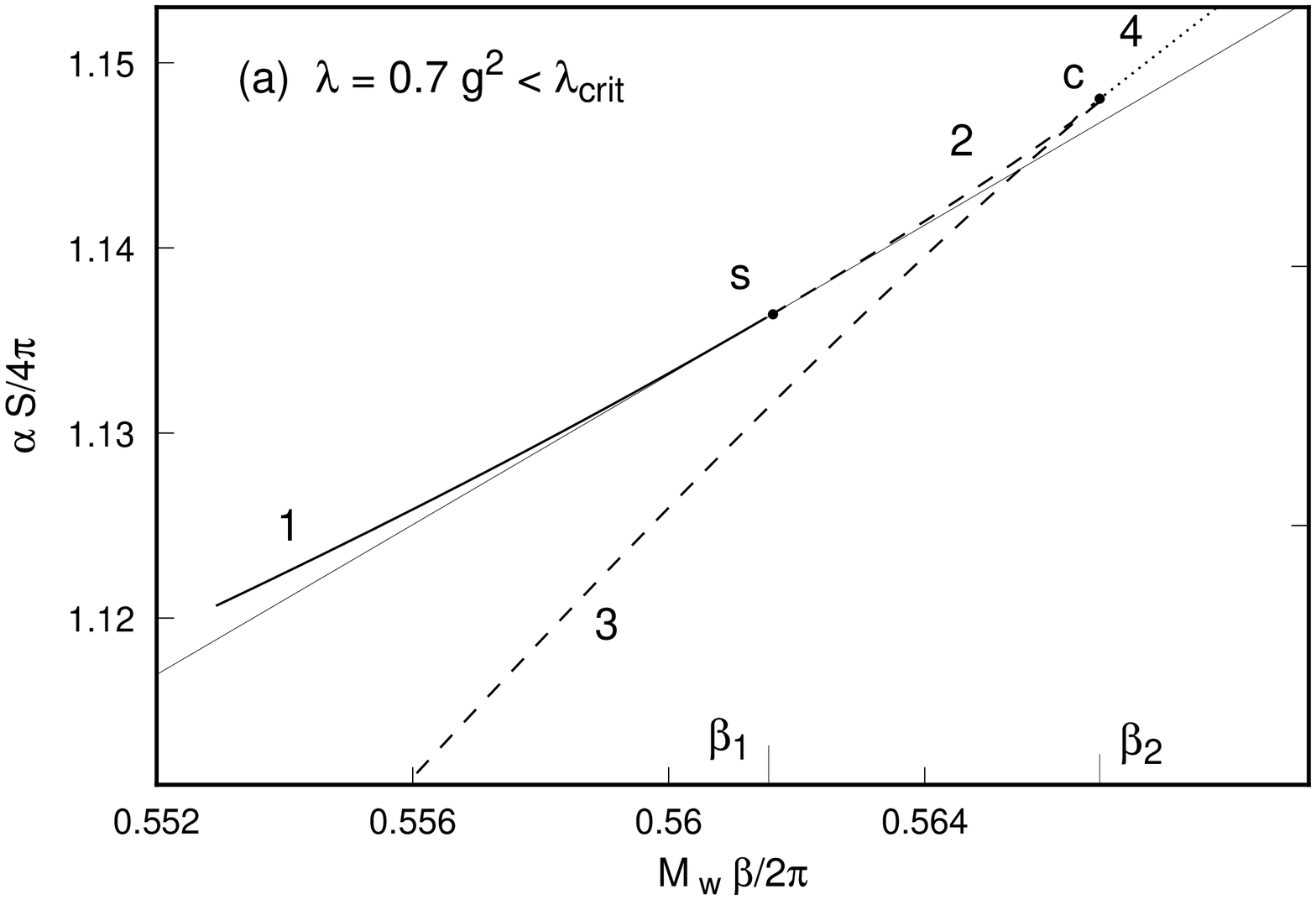}}
\epsfxsize=12cm
\centerline{\epsfbox{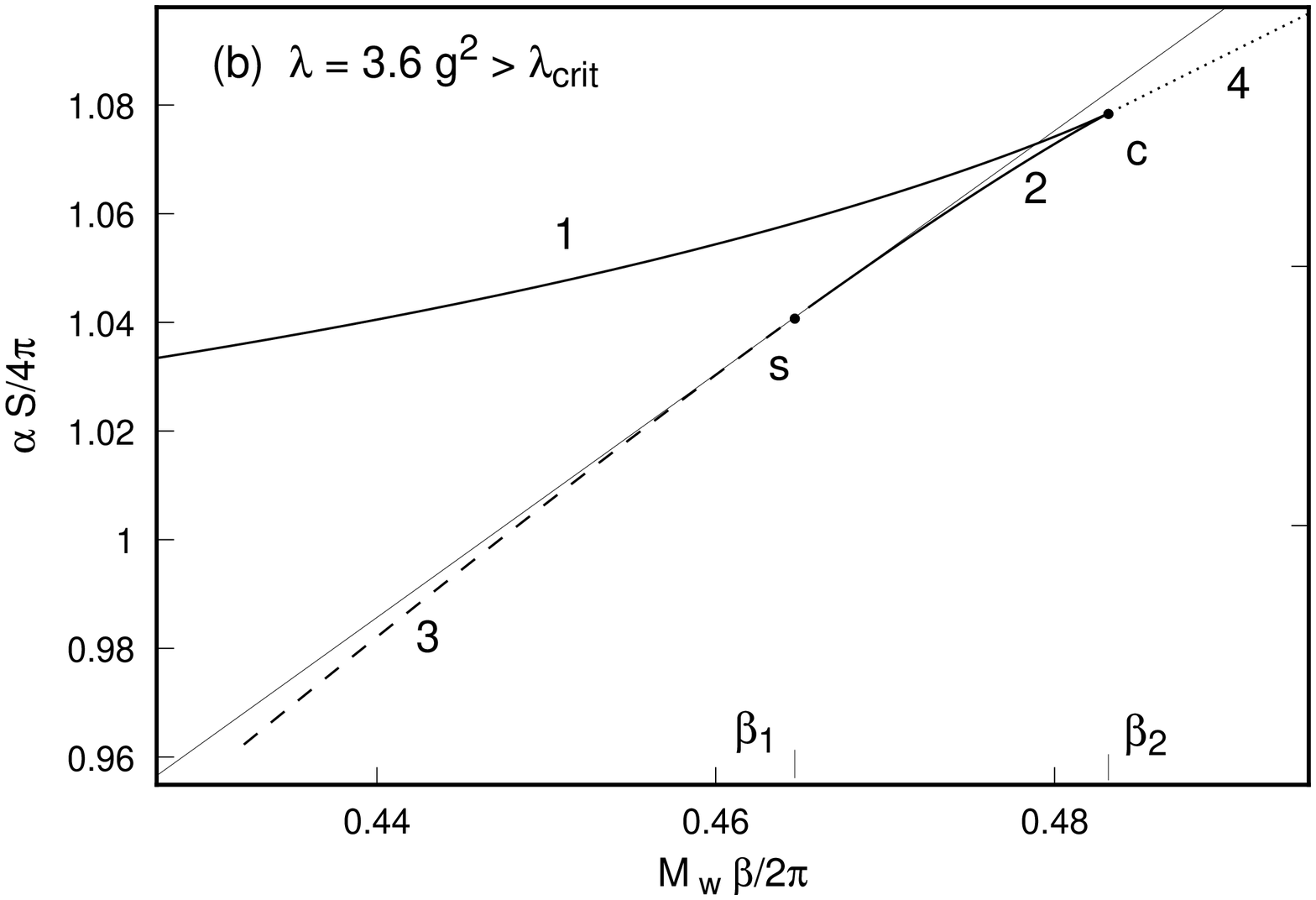}}
\vskip0.1cm
\caption {The action of the periodic instanton as a function of
period along the four distinct branches of solutions. The action
of the sphaleron is shown by the solid diagonal lines. In (a),
$\lambda=0.7 g^2 < \lambda_{\rm crit}$. The only real solutions
lie along branch 1, which joins onto the perturbative solutions
at small period. Branch 1 merges with the sphaleron at $s$, and may
be analytically continued onto branch 2. Branches 3 and
2 have real action and energy and form a cusp at point
$c$, beyond which along branch 4 the action becomes complex (only
the real part of the action is shown). The cusp $c$ always lies
above the sphaleron line in this regime of Higgs coupling. In
(b), $\lambda=3.6 g^2 > \lambda_{\rm crit}$ and the situation is
similar, except that it is now the complex branch 3 that merges
with the sphaleron. Branches 1 and 2 have real action and energy
and form a cusp $c$ that lies below the sphaleron line. Again,
the action and energy along branch 4 are complex.
}
\label{fig1}
\end{figure}
Since the numerical method and our code works equally well for
real or complex solutions, we were able to follow the action and
energy of the latter as well. We observe an interesting pattern 
of bifurcating solutions depending on the Higgs self-coupling $\lambda$,
as anticipated in ref. \cite{Kuz}.  

There are two regimes, separated by a critical value $\lambda_{\rm cr}$.
As illustrated in Fig.~1a, for $\lambda < \lambda_{\rm cr}$ the
perturbative solutions (branch 1) merge with the sphaleron
at some value $\beta=\beta_1$, which is similar in behavior
to the $O(3)$ sigma model. For $\beta>\beta_1$, branch 1 can be
analytically continued up to a second bifurcation point
$\beta_2$ to yield complex solutions whose energy and action
however remain real (\hbox{branch 2}). At $\beta_2$, branch 2
merges with yet another branch of complex solutions (branch 3),
and the action and energy of these solutions are also real. The
numerical results for this second complex branch suggest that
its action decreases monotonically with decreasing $\beta$, and
that at a sufficiently low (but positive) $\beta$ its action
{\it vanishes}, with potentially interesting physical
consequences (however, as indicated below, the energy along this
branch becomes arbitrarily large). Branches 2 and 3 form a cusp
$c$ at $\beta_2$, beyond which the solutions may be analytically
continued onto a fourth branch. The action and the
energy on branch 4 ($\beta>\beta_2$) are complex, and
the cusp $c$ always lies above the sphaleron line (only the real 
part of the action has been graphed).

Figure~1b illustrates the situation for $\lambda>\lambda_{\rm cr}$.
The pattern is similar, except in this case it is the complex
solutions along branch 3, and not the real solutions along branch
1, that merge with the sphaleron, and whose extension beyond
$\beta_1$ becomes branch 2. \hbox{Branch 1} (the perturbative
one) and
branch 2 are real up to $\beta_2$, where they form a cusp $c$
beyond which the solutions are complex with complex energy and
action (branch 4). In this case, the cusp always lies below the
sphaleron line, except at a critical coupling $\lambda_{\rm
crit}$ where the cusp intersects the sphaleron. We found
$\lambda_{\rm cr}\simeq 1.198 g^2$, corresponding to $M_H \simeq
3.096 M_W$, in good agreement with  $3.091 M_W$, obtained in
ref.~\cite{Frost} by a careful treatment of perturbations away
from the static sphaleron solution.

\begin{figure}[t]
\epsfxsize=12cm
\epsfysize=8cm
\centerline{\epsfbox{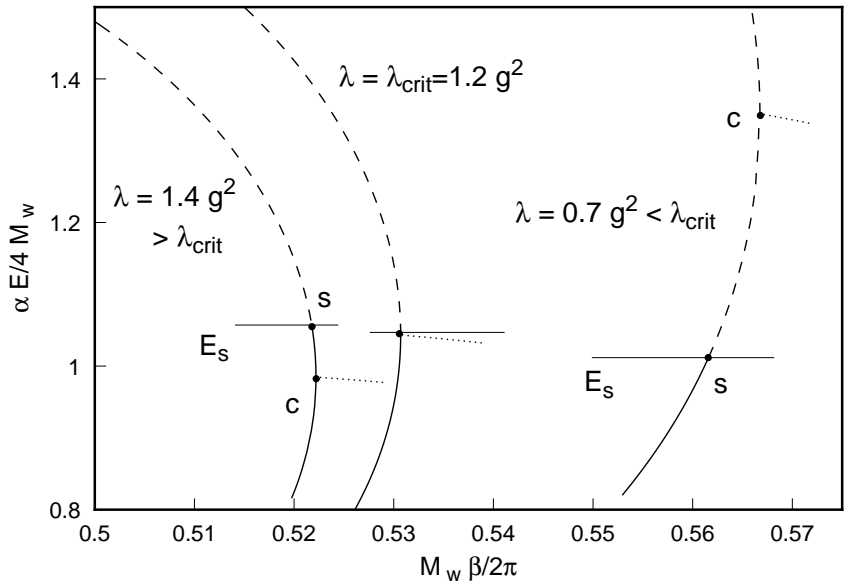}}
\noindent
\caption{The energy as a function of period near the bifurcation
points for two values of $\lambda$ on either side of the critical
Higgs coupling and for the critical coupling $\lambda_{\rm crit}$
itself. The solid horizontal lines labeled by $E_s$ denote the
constant sphaleron energy for each value of $\lambda$, while
the dashed, dotted, and solid curves are the energies of the
corresponding solutions in Figs.~1.
}
\label{fig2}
\end{figure}

The bifurcation structure for the energy of the complex solutions
is shown in Fig.~\ref{fig2} for two values of $\lambda$ on either
side of the critical value and for the critical coupling
$\lambda_{\rm crit}$ itself. The associated sphaleron energies
$E_s(\lambda)$ are illustrated by the short horizontal 
lines. The real solutions are indicated by the bold solid lines,
the complex solutions with real energy by the dashed lines, and
the complex solutions with complex energy by the dotted lines
extending down from the cusps $c$ (only the real part of the
energy has been graphed). The critical coupling $\lambda_{\rm
crit}$ occurs when the energy $E_0(\lambda)$ of the cusp crosses
the sphaleron line, {\it i.e.} when $E_0(\lambda_{\rm crit}) =
E_s(\lambda_{\rm crit})$.  Finally, it should be noted that the
equations are analytic (both in $\lambda$ and in $\beta$), and
therefore the total number of solutions cannot change at the
bifurcation points. Indeed, branches 1, 2 and 3 consist of two
independent solutions each (with the same energy and action),
whereas branch 4 consists of 4 solutions (in two conjugate
pairs). By continuity in $\lambda$, these (unsuspected) new
complex instanton solutions continue to exist even at lower
$\lambda$. The physical consequences of these new solutions for
rates of anomalous processes at finite energy in the electroweak
theory are currently under investigation.

\begin{figure}[t]
\epsfxsize=10cm
\centerline{\epsfbox{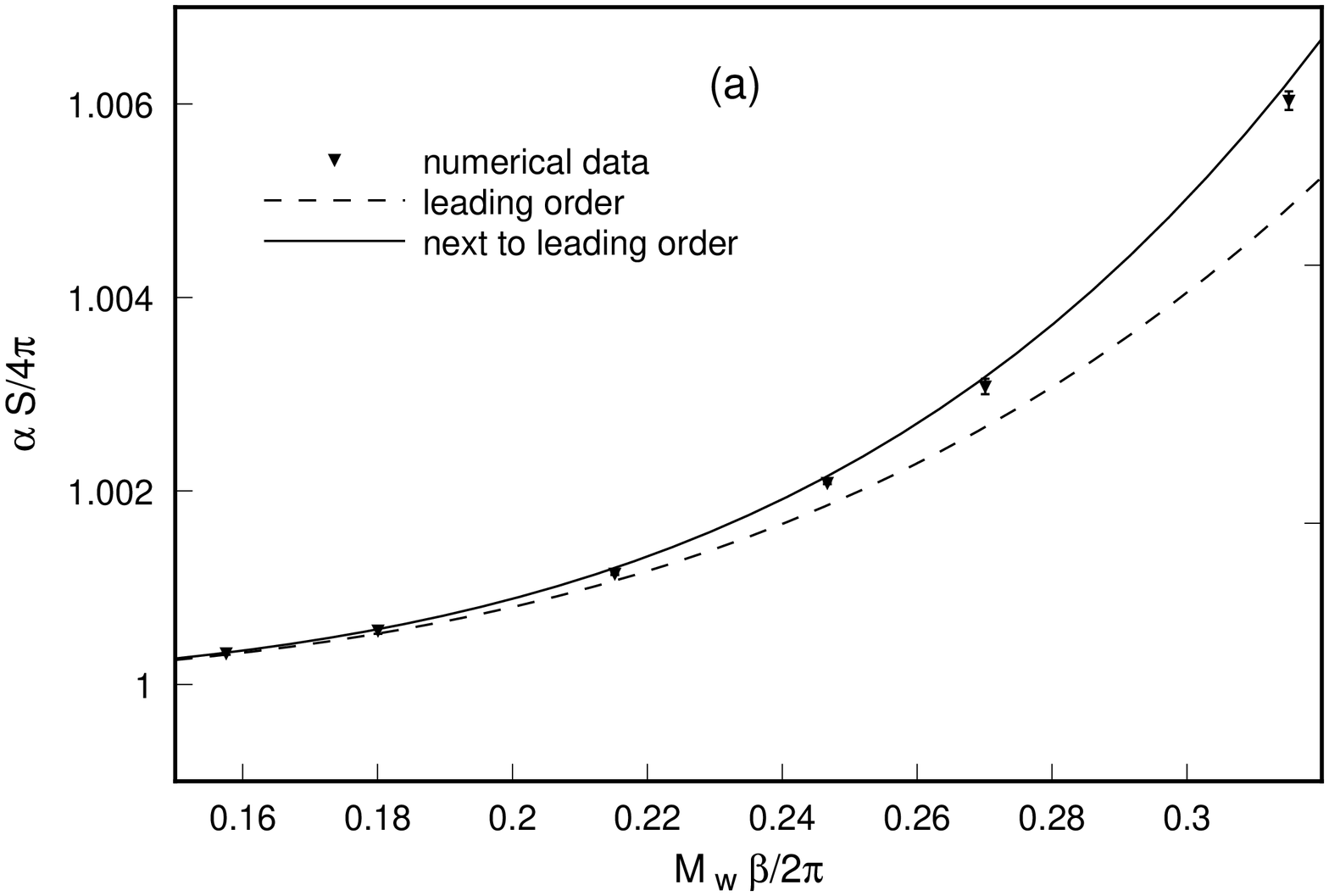}}
\epsfxsize=10cm
\centerline{\epsfbox{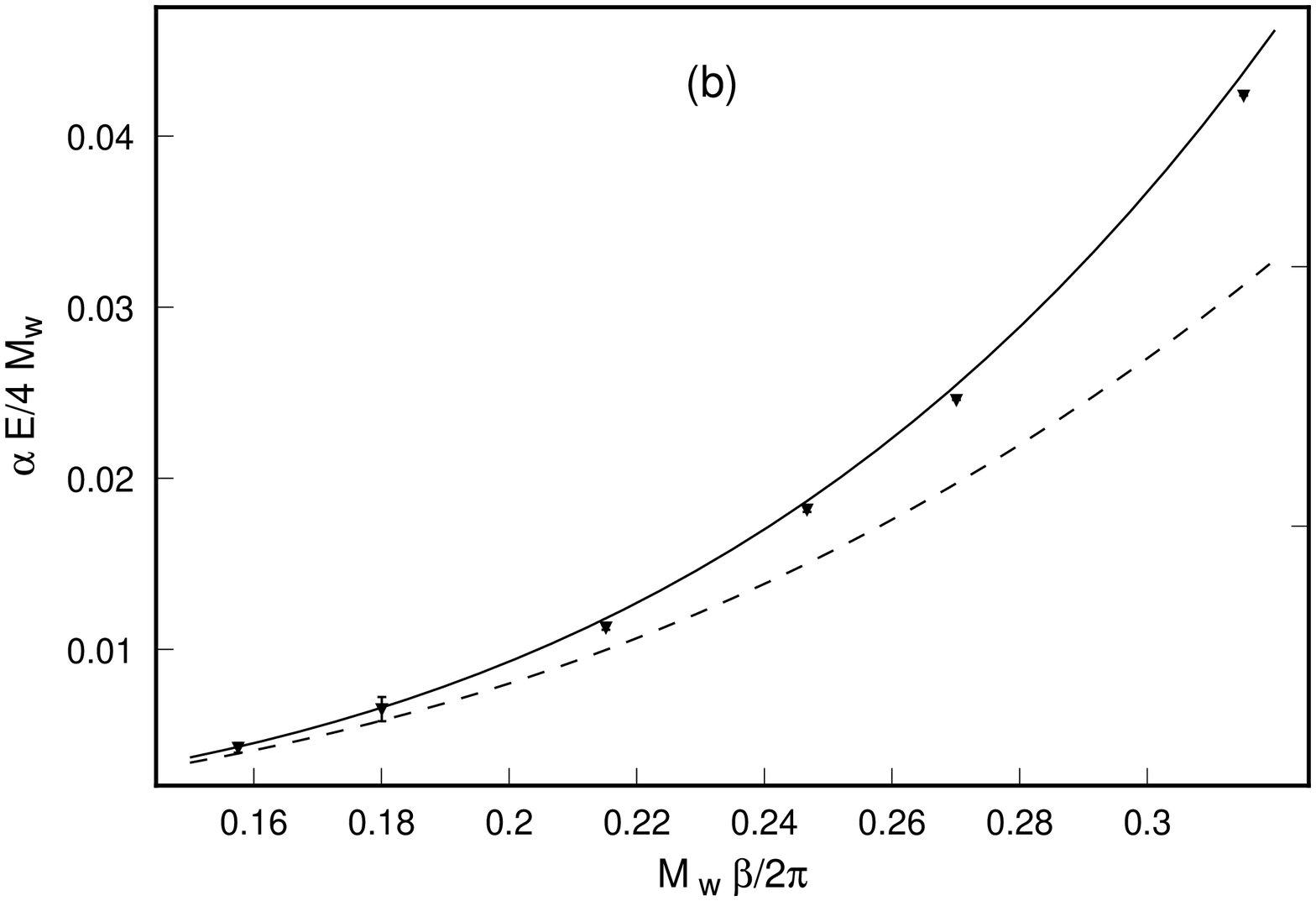}}
\noindent
\caption{The action (a) and energy (b) of the periodic instanton
as a function of $\beta$ in the perturbative regime (small
$\beta$). The numerical data points are denoted by the triangles,
the leading order calculations (14)-(15) by
the dashed lines, and the next to leading order calculations
(16)-(17) by the solid lines.
}
\label{fig3}
\end{figure}

To conclude, let us remark that the accuracy of our numerical
calculations is high enough that they can provide a verification
of the perturbative results for low energy, low period.
In Fig.~\ref{fig3} we plot our numerical results for action
and energy together with the results of the perturbative expansions
to leading order~(\ref{Spert})-(\ref{Epert}) and next to leading
order~(\ref{Spert2})-(\ref{Epert2}). The agreement with the next
to leading order perturbative results is quite good.

\vskip1.5cm
\noindent {\bf Acknowledgments}

We would like to acknowledge several helpful conversations with
V.~A.~Rubakov and L.~G.~Yaffe. This research was supported in
part under DOE grants DE-FG02-91ER40676 and DE-FG03-96ER40956,
the Russian Foundation for Basic Research grant 96-02-17449a,
and by the U.S.~Civilian Research and Development Foundation for
Independent States of FSU (CRDF) award RP1-187. One of us (P.~T.)
would like to acknowledge support from the Boston University
Center for Computational Science for two extended visits, during
which part of this work was completed.

\vskip0.5cm
\noindent {\bf Note Added in Proof:} Yaffe and Frost have
reported similar findings in ref.~\cite{Frosttwo}.

\end{document}